\documentclass[12pt]{article}
\usepackage{amsxtra,amssymb,amsthm,amsmath,latexsym,graphics}

\theoremstyle{plain}
\newtheorem{theorem}{Theorem}
\newtheorem{lemma}{Lemma}
\newtheorem{remark}{Remark}

\newcommand{\refT}[1]{Theorem~\ref{T:#1}}
\newcommand{\refS}[1]{Section~\ref{S:#1}}

\newcommand{\refL}[1]{Lemma~\ref{L:#1}}

\def\nd{\noindent}
\def\bysame{\rule{.5in}{.005in},\ }

\def\ve{{\varepsilon}}
\def\R{{\mathbb R}}

\def\C{{\mathbb C}}

\def\tildeD{\widetilde D}

\def\oH{{\overset{\circ}{H}}}
\def\oH1{{\overset{\circ}{H}\kern-.02in{}^1}}
\def\l{\ell}

\def\bee{\begin{equation*}}
\def\eee{\end{equation*}}
\def\be{\begin{equation}}
\def\ee{\end{equation}}
\def\d{\delta}

\begin{document}
\title{Computational method for acoustic wave focusing}

\author{A.G. Ramm\\
 Mathematics Department, Kansas State University, \\
 Manhattan, KS 66506-2602, USA\\
ramm@math.ksu.edu,\\ fax 785-532-0546, tel. 785-532-0580\\
and\\
S. Gutman\\
 Mathematics Department, University of Oklahoma, \\
 Norman, OK 73019, USA \\sgutman@ou.edu}


\date{}
\maketitle\thispagestyle{empty}

\begin{abstract}
Scattering properties of a material  are changed when the material
is injected with small acoustically soft particles. It is shown
that its new scattering behavior can be understood as a solution
of a potential scattering problem with the potential $q$
explicitly related to the density of the small particles. In this
paper we examine the inverse problem of designing a material with
the desired focusing properties. An algorithm for such a problem
is examined from the theoretical as well as from the numerical
perspective.
\end{abstract}

\footnote{MSC: 35J05, 35J10, 35R30, 74J25, 81U40, 81V05}
\footnote{PACS: 03.04.Kf}
\footnote{key words: "smart" materials, inverse scattering, scattering
of acoustic waves, computational method}


\section{Introduction}\label{S:1}
Let $D\subset\R^3$ be a bounded connected domain with
Lipschitz boundary $S$. Denote by $n_0(x)$
the refraction coefficient in $D$, $n_0(x)=1$ in
$D':=\R^3\setminus D$. Then the scattering of a plane acoustic  wave
$u_0=u_0(x)=e^{ik\alpha\cdot x}$, incident upon $D$,
is described by the system:
\be\label{e1}
[\nabla^2+k^2n_0(x)]u(x)=0\hbox{\quad in\quad} \R^3,
\ee
\be\label{e2} u(x)=u_0(x)+v(x),\ee
\be\label{e3}
 v(x)=A(\alpha',\alpha)\ \frac{e^{ikr}}{r}+ o\left(\frac{1}{r}\right),
 \qquad r:=|x|\to\infty, \quad \frac{x}{r}:=\alpha', \ee
where $v(x)$ is the scattered field, $\alpha\in S^2$ is the direction of the incident plane
wave, and $\alpha'$ is the direction of the scattered wave.
The coefficient $A(\alpha',\alpha)$ is called the scattering
amplitude, $k>0$ is the wave number, which is assumed to be fixed throughout
the paper. For this reason the dependence of $A$ on $k$ is not shown.

Let $D_m$, $1\leq m\leq M$, be a small particle, i.e.,
\be\label{e4}
  k_0a\ll 1, \hbox{\ where\ } a=\frac{1}{2}
  \max_{1\leq m\leq M} \,diam\, D_m,
\quad  k_0=k \max_{x\in D} |n_0(x)|. \ee
The geometrical shape of $D_m$ is arbitrary, but we assume that
each $D_m$ has a Lipschitz boundary. Moreover, the Lipschitz constant is the same
for every domain $D_m$. This
is a technical assumption which can be relaxed. It allows one
to use the properties of the electrostatic potentials. Let
\be\label{e5}  d:=\min_{m\not= j}dist(D_m,D_j). \ee
Assume that
\be\label{e6} a\ll d. \ee
 We do not assume that  $d\gg \lambda_0$, that is, that the distance between
the particles is much larger than the wavelength. Under our assumptions it is possible that there are
many small particles within the distances of the order of magnitude of the wavelength.

The particles are assumed to be acoustically soft, i.e.,
\be\label{e7}
 u|_{S_m}=0\qquad 1\leq m\leq M.
 \ee
As a result of the distribution of many small particles in $D$,
one obtains a new material. We would like this "smart" material to
have some desired properties.
Specifically, we want this material to scatter the incident plane wave
according to an a priori given desired radiation pattern, for example to
focus the incident wave within a given solid angle.
 Is this possible? If yes, then how does one distribute the small particles
in order to create such a material?
In mathematical terms the problem is

\textit{
Given an arbitrary function $f(\beta)\in L^2(S^2)$, can one
distribute small particles in $D$ so that the resulting medium
generates the radiation pattern $A(\beta):= A(\beta,\alpha)$, at
a fixed $k>0$ and a fixed $\alpha\in S^2$, such that
\be\label{e8}
 \|f(\beta)-A(\beta)\|_{L^2(S^2)}\leq\ve,
\ee
where $\ve>0$ is an arbitrary small fixed number?
}

The answer is yes.
It is contained in the following Theorem.
\begin{theorem}\label{T:1}
For any $f\in L^2(S^2)$, an arbitrary
small $\ve>0$, any fixed
$\alpha=\alpha_0\in S^2$, any fixed $k=k_0>0$,
and any bounded domain $D\subset \R^3$,
there exists a (non-unique) potential $q(x)\in L^2(D)$, such that
\eqref{e8}
holds. \end{theorem}

The relation between the particle distribution density and the potential
$q$ is explained in \refS{2} and \refS{3}. In \refS{4} we give an algorithm for calculating
such a potential. Numerical results are presented in \refS{5}.
 Our solution of this problem is based on our earlier results on wave scattering
by small bodies of arbitrary shapes, see Ramm (2005b), as well as Ramm
(2006a,b).

\section{Scattering by many small particles}\label{S:2}

If many small particles
$D_m$, $1\leq m\leq M$, are embedded in $D$, $u\mid_{S_m}=0$, where $S_m$
is the boundary of $D_m$, then
the scattering problem is:
\be\label{2:e3}
 [\nabla^2+k^2-q_0(x)]u=0\hbox{\ in\ }\R^3\setminus \bigcup^M_{m=1}D_m,
\ee
\be\label{2:e4}  u\mid_{S_m}=0, \quad m=1,\dots,M,\ee
\be\label{2:e5}
 u(x)=u_0(x)+A(\beta,\alpha)\,\frac{e^{ikr}}{r}+o\left(\frac{1}{r}\right),
 \quad r=|x|\to\infty,\quad \beta=\frac{x}{r},\ee
and the solution $u(x)$ is called the scattering solution. Here
$n_0(x)$ is a given refraction coefficient, $n_0(x)>0$ in $\R^3$,
$u$ is the acoustic pressure,
\[q_0(x) :=k^2[1-n_0(x)]=0 \hbox{\quad  in \quad} D'.\]

Under the assumptions of \refS{1}, we can have $d\ll \lambda$. We also
assume that the quantity
$a/d^3$ has a finite non-zero limit as $M\to\infty$ and
$a/d\to 0$.
More precisely, if $C_m$ is the electrical capacitance of the conductor
with the shape $D_m$, then we assume the existence of a limiting density $C(x)$
of the capacitance per unit volume around every point $x\in D$:
\be\label{2:e7}
 \lim_{M\to\infty}\sum_{D_m\subset\tildeD} C_m=\int_{\tildeD} C(x)dx, \ee
where $\tildeD\subset D$ is an arbitrary subdomain of $D$. Note that the
density of the volume of the small particles per unit volume is
\[
O\left(\frac{a^3}{d^3}\right)\to 0\quad \text{as}\quad\frac{a}{d}\to 0.
\]
One can prove (see Ramm (2005b), p.103) that in the limit $M\to\infty$ the
function $u$ solves the equation
\be\label{2:e8}
 [\nabla^2+k^2-q(x)]u=0 \hbox{\ in\ }\R^3,\quad q(x)=q_0(x)+C(x),\ee
where $C(x)$ is defined in \eqref{2:e7}, and $A(\beta,\alpha)$ in \eqref{2:e5}
corresponds to the potential $q(x)$   (see also Marchenko and Kruslov
(1974), where similar
homogenization-type problems are discussed).

If all the small particles are identical, $C_0$ is the capacitance of a
conductor in the shape of a particle, and $N(x)$ is the number of small
particles per unit volume around point $x$, then, up to the quantity of
higher order of smallness as $\frac a d\to 0$, we have:
\[C(x)=N(x)C_0.\]
Therefore,
\[q(x)=q_0(x)+N(x)C_0,\]
and
\be\label{2:e9}  N(x)=\frac{q(x)-q_0(x)}{C_0}. \ee
Thus, one has an explicit one-to-one correspondence between $q(x)$ and the
density $N(x)$ of the embedded particles per unit volume.

\begin{remark} If the boundary condition on $S_m$ is of impedance type:
\[  u_N=\zeta u\hbox{\quad on\quad} S_m, \]
where $N$ is the exterior unit normal to the boundary $S_m$, and $\zeta$ is
a complex constant, the impedance, then the capacitance $C_0$
in formula (9) should be replaced by
\[ C_\zeta =\frac{C_0}{1+\frac{C_0}{ \zeta |S|}},\]
where $|S|$ is the surface area of $S$, and the corresponding potential
$q(x)$ will be complex-valued, see Ramm (2005b), p. 97.
\end{remark}

\section{Scattering solutions}\label{S:3}

To establish  \refT{1}, recall that for a fixed $k>0$  the scattering problem
\eqref{e1}--\eqref{e3} is equivalent to the Schro\"dinger
scattering problem for the potential $q(x)$:

\be\label{e13}
u_q=u_0-\int_D g(x,y) q(y)u_q(y)dy,
  \quad g(x,y):=\frac{e^{ik|x-y|}}{4\pi|x-y|}, \ee
for which the scattering solution $u= u_q$ is the unique solution.
The corresponding scattering amplitude is
\be\label{e14}
  A(\alpha',\alpha)=-\frac{1}{4\pi}
  \int_D e^{-ik\alpha'\cdot x} q(x)u_q(x,\alpha)dx,\ee
where the dependence on $k$ is dropped since $k>0$ is fixed.

If $q$ is known, then $A:=A_q$ is known.
Let $q\in L^2(D)$ be a potential and $A_q(\alpha',\alpha)$ be
the corresponding scattering amplitude. Fix $\alpha\in S^2$ and
denote
\be\label{e15}
  A(\beta):=A_q(\alpha',\alpha), \qquad \alpha'=\beta. \ee
Then
\be\label{e16}
  A(\beta)=-\frac{1}{4\pi}\int_D e^{-ik\beta\cdot x}h(x)dx,
  \qquad h(x):=q(x)u_q(x,\alpha). \ee

Our goal in \refT{1} is to find a potential $q$ for which \eqref{e8}
is satisfied. First, we find an
 $h(x)$  that satisfies
\be\label{e12}
 \|f(\beta)+\frac{1}{4\pi} \int_D e^{-ik\beta\cdot x}
 h(x)dx\|_{L^2(S^2)} <\ve. \ee
The existence of such an $h$ follows from the following Theorem.
\begin{theorem}\label{T:2}
Let $f(\beta)\in L^2(S^2)$ be arbitrary. Then
\be\label{e17}
  \inf_{h\in L^2(D)} \bigg\| f(\beta)+
  \frac{1}{4\pi} \int_D e^{-ik\beta\cdot x}h(x)dx
  \bigg\|_{L^2(S^2)}  =0.\ee
\end{theorem}

\begin{proof}[Proof of \refT{2}]
If \eqref{e17} fails, then there is a function $f(\beta)\in
L^2(S^2)$, $f\not=0$, such that
\be\label{e18}
  \int_{S^2} d\beta f(\beta) \int_D e^{-ik\beta\cdot x}
  h(x)dx=0 \qquad \forall h\in L^2(D).\ee
This implies
\be\label{e19}
  \varphi(x):=\int_{S^2} d\beta f(\beta)e^{-ik\beta\cdot x}=0
  \qquad \forall x\in D.\ee
The function $\varphi(x)$ is an entire function of $x$.
Therefore \eqref{e19} implies
\be\label{e20}
  \varphi(x)=0\qquad \forall x\in \R^3.\ee
This and the injectivity of the Fourier transform imply
$f(\beta)=0$. Note that $\varphi(x)$ is the Fourier transform
of the distribution $f(\beta)\delta(k-\lambda)\lambda^{-2}$,
where $\delta(k-\lambda)$ is the delta-function and
$\lambda\beta$ is the Fourier transform variable.
The injectivity of the Fourier transform implies
$f(\beta)\lambda^{-2}\delta(k-\lambda)=0$, so $f(\beta)=0$.
 \refT{2} is proved.
\end{proof}

To find an $h$ that satisfies \eqref{e12} one can proceed as follows.
Let $\{Y_\l(\beta)\}_{\l=0}^{\infty}$, $Y_\l=Y_{\l,m}$, $-\l\leq
m\leq\l$, be the orthonormal in $L^2(S^2)$ spherical harmonics,
\be\label{e21}
  Y_{\l,m}(-\beta)=(-1)^\l Y_{\l,m}(\beta), \quad
\overline{Y_{\l,m}(\beta)}=(-1)^{\l+m}Y_{\l,m}(\beta), \ee
\be\label{e22}
  j_\l(r):=\left(\frac{\pi}{2r}\right)^{1/2}
  J_{\l+\frac{1}{2}}(r), \ee
where $J_\l$ are the Bessel functions and the overbar stands for the
complex conjugate. It is known that
\be\label{e23}
  e^{-ik\beta\cdot x}=\sum_{\l=0,-\l\leq m\leq\l}
  4\pi(-i)^\l j_\l(kr) \overline{Y_{\l,m}(x^0)}
  Y_{\l,m}(\beta), \quad x^0:=\frac{x}{|x|}. \ee
Let us expand $f$ into the Fourier series with respect to spherical harmonics:
\be\label{e24}
  f(\beta)=\sum_{\l=0,-\l\leq m\leq\l} f_{\l,m}
Y_{\l,m}(\beta).\ee
Choose $L=L(\ve)$ such that
\be\label{e25}
  \sum_{\l>L}|f_{\l,m}|^2\leq \ve^2. \ee
With so fixed $L$, take $h_{\l,m}(r)$, $0\leq\l\leq L$,
$-\l\leq m\leq\l$, such that
\be\label{e26}
  f_{\l,m}=-(-i)^\l \left(\frac{\pi}{2k}\right)^{1/2}
  \int^b_0 r^{3/2} J_{\l+\frac12}(kr) h_{\l,m}(r)dr,\ee
where $b>0$, the origin $O$ is inside $D$, the ball centered at the
origin
and of radius $b$ belongs to $D$, and $h_{\l,m}(r)=0$ for $r>b$.
There are many choices of $h_{\l,m}(r)$ which satisfy
\eqref{e26}. If \eqref{e25} and \eqref{e26} hold, then the norm
on the left-hand side of \eqref{e17} is smaller than $\ve$.

A possible explicit choice of $h_{\l,m}(r)$ is
\be\label{e27}
 h_{\l,m}=\left\{\begin{array}{ll}-(-i)^\l
 \frac{f_{\l,m}}
      {\sqrt{\frac{\pi}{2k}} g_{1,\l+\frac12}(k)}, & \l\leq L,\\
 0\hfill, & \l>L\end{array}\right. \ee
where $g_{\mu,\nu}(k):=\int^1_0 x^{\mu+\frac12}J_\nu(kx)dx$. This integral
can be calculated analytically, see Bateman and  Erdelyi (1954), formula 8.5.8. We
have assumed that $h(x)=0$ for $|x|>1$, and $b=1$ in \eqref{e26}.
Finally, let
\be\label{e28}
h(x)=\sum^L_{\l=0}h_{\l,m}(r) Y_{\l,m}(\alpha').
\ee
This function satisfies inequality (19) by the construction.

\section{Reconstruction of the potential}\label{S:4}
In the previous section we have shown how to find a function $h\in L^2(D)$
that satisfies \eqref{e12}. In this section a potential $q$
satisfying the conditions of \refT{1} is constructed from such an $h$.
The possibility of such a reconstruction follows from the following
result.
\begin{theorem}\label{T:3}
Let $h\in L^2(D)$ be arbitrary. Then
\be\label{e29}
  \inf_{q\in L^2(D)} \|h-qu_q(x,\alpha)\|=0.\ee
Here $\alpha\in S^2$ and $k>0$ are arbitrary, fixed. Moreover, if
$||h||_{L^2(D)}$ is sufficiently small, then
 there exists a potential $q$ such that
\be\label{e30}
h(x)=q(x)u_q(x,\alpha).
\ee
\end{theorem}

This Theorem follows from \refL{3} and \refL{4} stated and proved below.
For convenience let us summarize the method for finding a potential $q$
satisfying the conditions of \refT{1}.

{\bf Method for Potential Reconstruction.}

Let $\ve>0$.

\nd\underbar{Step 1.} Given an arbitrary function $f(\beta)\in L^2(S^2)$
find $h\in L^2(D)$ such that \eqref{e12} holds. This can be done using
\eqref{e27}. Let
\be\label{4:e2}
h(x)=\sum^L_{\l=0}h_{\l,m}(r) Y_{\l,m}(\alpha'),
\ee
where $L=L(\ve)$, see (30).

\nd\underbar{Step 2.} Use $h$, obtained in Step 1, to find a potential
$q\in L^2(D)$ satisfying \[ \|h-qu_q(x,\alpha)\|<\ve. \]
 For $f$ with a sufficiently small norm $||f(\beta)||_{ L^2(S^2)}$
 such a potential $q$ can be found using formula \eqref{4:e9}, see below.
Formula \eqref{4:e9} can be used for any $f$ for which condition
\eqref{4:e8} holds.

\nd\underbar{Step 3.} This potential $q$ generates the scattering
amplitude $A(\beta)$ at fixed $\alpha$ and $k$, such that \[
 \|f(\beta)-A_q(\beta)\|_{L^2(S^2)}\leq C\ve
\]
 holds for some constant $C$, independent of $\ve$.

Indeed, let $\|\cdot\|:=\|\cdot\|_{L^2(S^2)}$. Then
\be\label{4:e15}\begin{aligned}
 \|f(\beta)-A_q(\beta)\|
 &=\|f(\beta)+\frac{1}{4\pi}\int_D e^{-ik\beta\cdot x} q(x)u(x)dx\| \\
 &\leq\|f(\beta) +\frac{1}{4\pi}\int_D e^{ik\beta\cdot x}hdx\| +\frac{\ve|D|}{4\pi}\\
 &\leq \ve+\frac{\ve|D|}{4\pi},
 \quad |D|=meas\,D. \end{aligned}\ee
This  concludes the proof of \refT{1}.

\begin{lemma}\label{L:3}
Assume that $\sup_{x\in D}\bigg| \int_D ghdy\bigg|<1$, or, more generally,
that
\be\label{4:e8}
 \sup_{x\in D}\bigg|u_0(x)- \int_D g(x,y)h(y)dy\bigg|>0,\qquad
 g=g(x,y):=\frac{e^{ik|x-y|}}{4\pi|x-y|}. \ee
Then equation \eqref{e30} has a unique solution:
\be\label{4:e9}
 q(x)=\frac{h(x)}{u_0(x)-\int_D g(x,y)h(y)dy}, \qquad
 u_0(x)=e^{ik\alpha\cdot x}, \quad q\in L^2(D). \ee
\end{lemma}

\begin{remark} It follows from \refT{2} and the discussion afterward
 that $f$ and $h$ are proportional, so that if
$||f||_{L^2(S^2)}$ is sufficiently small, then $||h||_{L^2(D)}$ is small,
and then
condition  \eqref{4:e8}  is satisfied.
\end{remark}

\begin{proof}[Proof of \refL{3}]
The scattering solution corresponding to a potential $q$ solves the
equation
\be\label{4:e13}
 u=u_0-\int_D g(x,y)q(y)u(y)dy, \qquad u_0:=e^{ik\alpha\cdot x}. \ee If
$h(x)=q(x)u_q(x,\alpha)$ holds, i.e., if $h$ corresponds to a $q\in
L^2(D)$, then $u=u_0-\int_D ghdy$. Multiply this equation by $q$ and get
\[q(x)u(x)=q(x)u_0(x)-q(x)\int_D g(x,y)h(y)dy.\] Using \eqref{e30} and
solving for $q$, one gets \eqref{4:e9}, provided that \eqref{4:e8} holds.
Condition \eqref{4:e8} holds if $\|h\|_{L^2(D)}$ is sufficiently small.
One has \be\label{4:e14}
 \bigg|\int_D g(x,y)hdy\bigg|\leq \frac{1}{4\pi} \sup_x
\bigg\|\frac{1}{|x-y|}\bigg\|_{L^2(D)} \|h\|_{L^2(D)},\ee
and
\[
\frac{1}{4\pi}\left(\int_D \frac{dy}{|x-y|^2}\right)^{\frac12}\leq
 \frac{\sqrt{a}}{\sqrt{4\pi}},
\]
 where $a=0.5 diam D$. If, for
example,
\[
\frac{\sqrt{a}}{\sqrt{4\pi}}\|h\|_{L^2(D)}<1,
\]
 then condition \eqref{4:e8}
holds, and formula \eqref{4:e9} yields the corresponding potential.
This explains the role of the "smallness" assumption.
\end{proof}

\begin{remark}\label{R:4}
If \eqref{4:e8} fail, then formula \eqref{4:e9} may yield a
$q\notin L^2(D)$. As long as formula \eqref{4:e9} yields a potential
$q\in L^p(D),\,\, p\geq 1,$ our arguments essentially remain valid.
In our presentation we have used $p=2$ because the numerical
minimization in $L^2$-norm is simpler.

The difficulty arises when formula \eqref{4:e9} yields a
potential which is not locally integrable. Numerical
experiments showed that this case did not occur in practice
in several test examples in which the "smallness" condition
was not satisfied.

We prove that a suitable small perturbation $h_\delta$
of $h$ in
$L^2(D)$-norm yields by formula \eqref{4:e9} a bounded
potential $q_\delta$.
This means that the "smallness"
restriction on the norm of $f$ is not essential.
\end{remark}

\begin{lemma}\label{L:4} Assume that $h$ is analytic in $D$
and bounded in the closure of $D$. There exists a small
perturbation $h_\d$ of $h$,
$||h-h_\d||_{L^2(D)}<\d$, such that the function
$$q_\d:=\frac{h_\d(x)}{u_0(x)-\int_Dg(x,y)h_\d(y)dy}$$ is
bounded.
\end{lemma}

{\it Outline of proof}.
Suppose that for a given $h\in L^2(D)$ condition \eqref{4:e8}
is not satisfied. Let us  approximate $h$ by an analytic
function $h_1$ in $D$, for example, by a polynomial, so that
  $$||f(\beta)+\frac{1}{4\pi}\int_D e^{-ik\beta\cdot
x} h_1(x)dx||<\ve.$$
Denoting $h_1$ by $h$ again, we may assume that $h$ is
analytic in $D$ and in a domain which contains $D$.
We prove that it is possible
to perturb $h$ slightly so that for the perturbed $h$,
denoted $h_\d$, condition \eqref{4:e8}
is satisfied, and formula \eqref{4:e9} yields a potential
$q_\d \in L^2(D)$, for  which inequality \eqref{e8}
holds, see Ramm (2006c) for details.

Finally we make some remarks about ill-posedness of our algorithm
for finding $q$ given $f$. This problem is ill-posed because an
arbitrary $f\in L^2(S^2)$ cannot be the scattering amplitude
$A_q(\beta)$ corresponding to a compactly supported potential $q$.
Indeed, it is proved in Ramm (1992), Ramm (2002), that 
$A(\beta)$
is infinitely differentiable on $S^2$ and is a restriction to
$S^2$ of a function analytic on the algebraic variety in $\C^3$,
defined by the equation $\beta\cdot\beta=k^2$. Finding $h$
satisfying \eqref{e12} is an ill-posed problem if $\ve$ is small.
It is similar to solving the first-kind Fredholm integral equation
$$\frac{1}{4\pi}\int_D e^{-ik\beta\cdot
x}h(x)dx=-f(\beta)$$
whose kernel is infinitely smooth. Our solution \eqref{e27} shows
the ill-posedness of the problem because the denominator in
\eqref{e27} tends to zero as $\ell$ grows. Methods for stable
solutions of ill-posed problems (see Ramm (2005a))
 should be
applied to finding $h$. If $h$ is found, then $q$ is found
by formula \eqref{4:e9}, provided that \eqref{4:e8} holds. If
\eqref{4:e8} does not hold, one perturbs slightly $h$
according to \refL{4}, and get a potential $q_\d$ by formula
\eqref{4:e9} with $h_\d$ in place of $h$.

\section{Numerical results}\label{S:5}

In this section we present results of numerical experiments
for a design of the material capable of focusing the
incoming plane wave into a desired solid angle. First, let
us note that a direct implementation of the algorithm
presented in the previous sections produces potentials $q$
with large magnitudes (of the order of $10^5$) in our 
examples. This happens because of the ill-posedness of 
the inverse scattering problem.
To remedy this situation we have introduced an additional 
step
in the potential reconstruction algorithm stated in
\refS{4}.

\nd\underbar{Step 1b.} Let $h_{l,m}$ be the coefficients of $h$ obtained
according to \eqref{e27}.  Bound the magnitudes of the coefficients by a predetermined
constant $T>0$, that is 
\[
 h'_{l,m}= \begin{cases} h_{l,m}\quad & \text{if}
\quad |h_{l,m}|\leq T\\ 
\frac{h_{l,m}}{|h_{l,m}|}T\quad & \text{if} \quad |h_{l,m}|>T. \end{cases}
\]
 Let \be\label{5:e1} h(x)=\sum^L_{\l=0}h'_{\l,m}(r) Y_{\l,m}(\alpha').
\ee

The bound on the function $h$ has the effect of bounding the potential
$q$. This procedure regularizes the ill-posedness of the
reconstruction process as discussed at the end of \refS{4}. The
ill-posedness manifests itself in the divergence of the series
\eqref{4:e2} with $L=\infty$, when the regularization
we have used by introducing the bounding constant, is not applied. 
However, from the 
numerical observations, the series
\eqref{5:e1} practically did not change with the increase in $L$ for
$L>5$. As expected, an increase in the value of $T$ improves the
precision of the approximation of the desired scattering amplitude $f$,
but it also increases the magnitude of the potential $q$. A reduction in
the value of $T$ leads to a deteriorating approximation.

In all the experiments the incident direction $\alpha=(0,0,1)$,
$k=1.0$ and $L=6$. The domain $D$ is the ball of radius $1$
centered in the origin. In our first numerical experiment the goal
was to focus the incoming plane wave into the solid angle
$0\leq\theta\leq \pi/4$, where $\theta$ is the polar angle
measured from the incident direction $\alpha=(0,0,1)$. Figure 1
shows the cross-section through the incident direction of the
desired (dotted line) and the attained absolute value (solid line)
of the scattering amplitude.

Figure 2 shows the contour plot of the absolute value of the
recovered potential $q$ in a cross-section through the $z$-axis.
The darker colors correspond to the larger values of $|q|$. In
this experiment the maximum of the absolute value of the potential
$q$ was about $1290$ corresponding to the bounding constant
$T=100$. This value of $T$ was found by examining numerical results with
larger and smaller values for this constant. For smaller $T$ the
resulting radiation pattern has smaller magnitudes, i.e. the plot of
its absolute value is
located closer to the origin. For larger values of $T$ the maximal value $|q|$ of the
potential approaches the order of $10^5$.

Similarly, Figures 3 and 4 show the results of the numerical experiment
aimed at focusing the same incident plane wave into the solid angle
$0.2\pi\leq\theta\leq 0.5\pi$. The maximum of $|q|$ was about
$1840$ in this case, corresponding to  the bounding constant $T=800$.
This value of $T$ was found experimentally as above. For
smaller values of $T$ the resulting radiation pattern has a significant
component in the region $|\theta|\leq 0.2\pi$, i.e. it produces a poor
approximation for the desired scattering amplitude.

\begin{figure}
\begin{center}
\includegraphics*{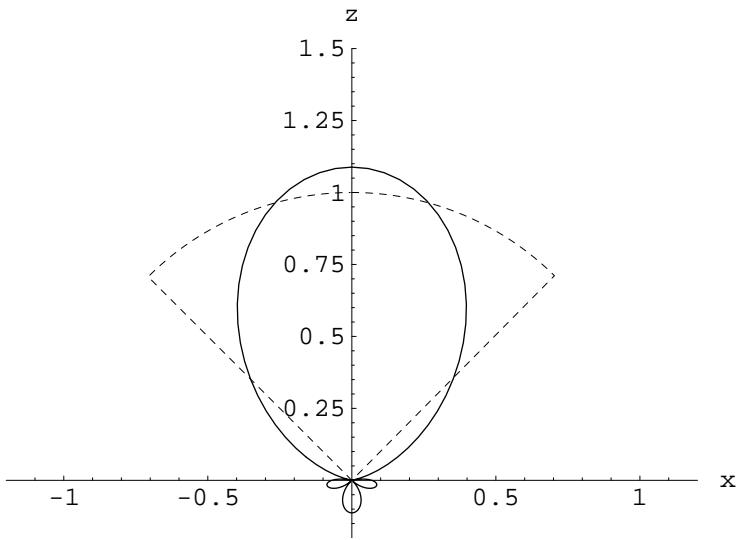}
\end{center}
\caption{Attained (solid line), and targeted (dotted line) scattering
amplitude $f(\beta)$ in experiment 1.}
\end{figure}

\begin{figure}

\begin{center}
\includegraphics*{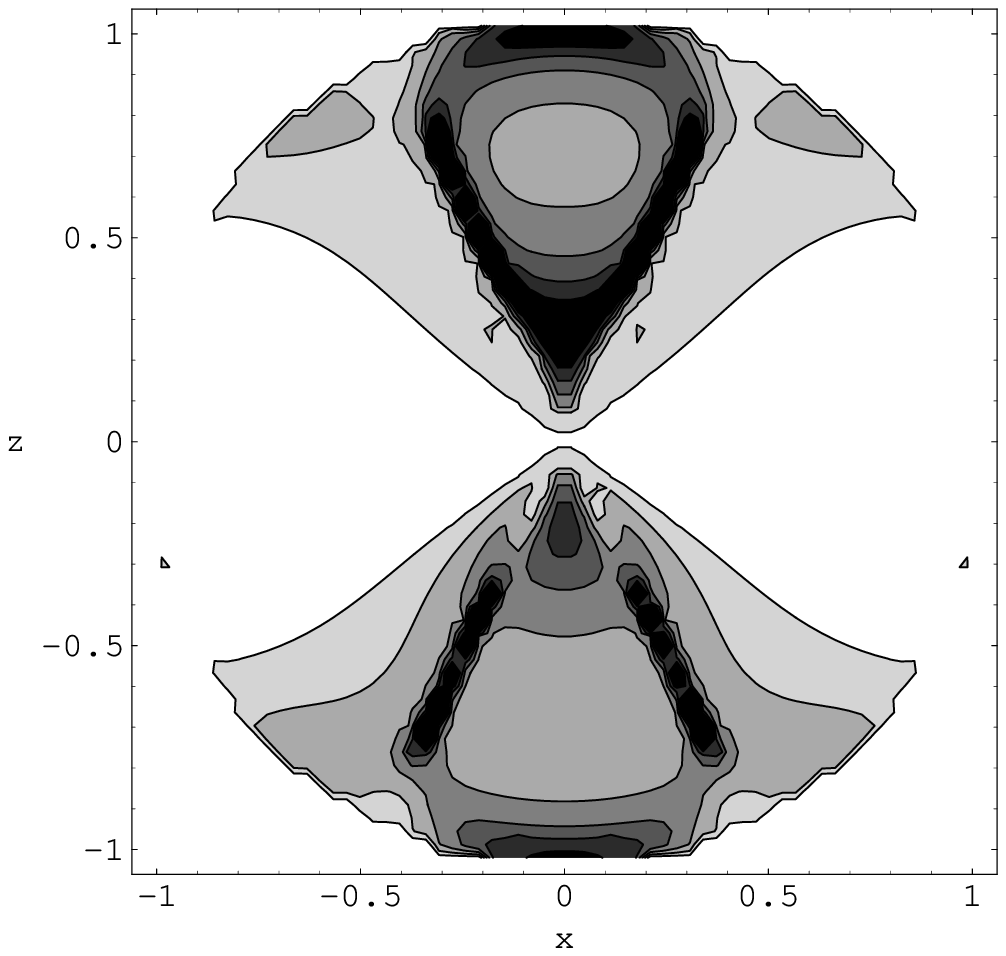}
\end{center}
\caption{Contour plot of the potential $q$ in experiment 1.}
\end{figure}

\begin{figure}
\begin{center}
\includegraphics*{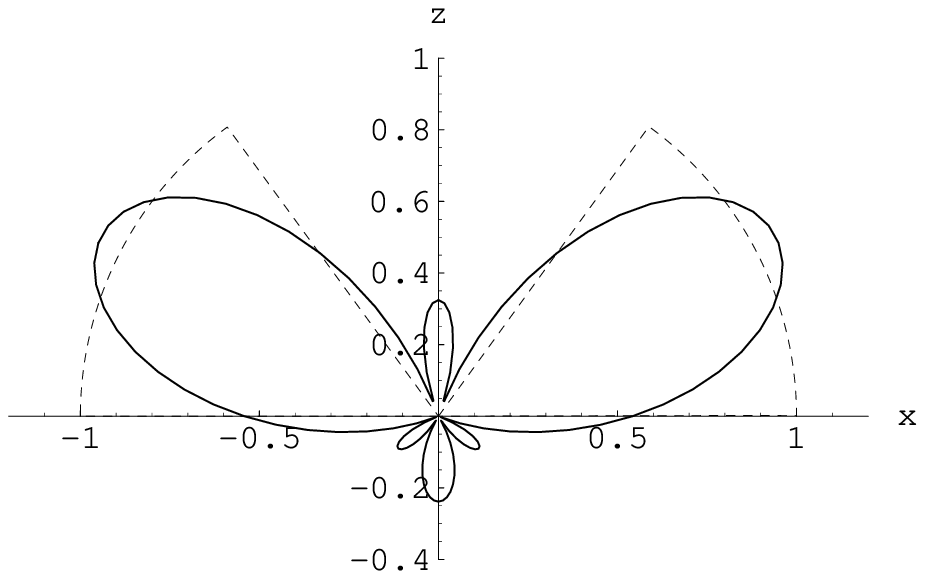}
\end{center}
\caption{Attained (solid line), and targeted (dotted line) scattering
amplitude $f(\beta)$ in experiment 2.}
\end{figure}

\begin{figure}

\begin{center}
\includegraphics*{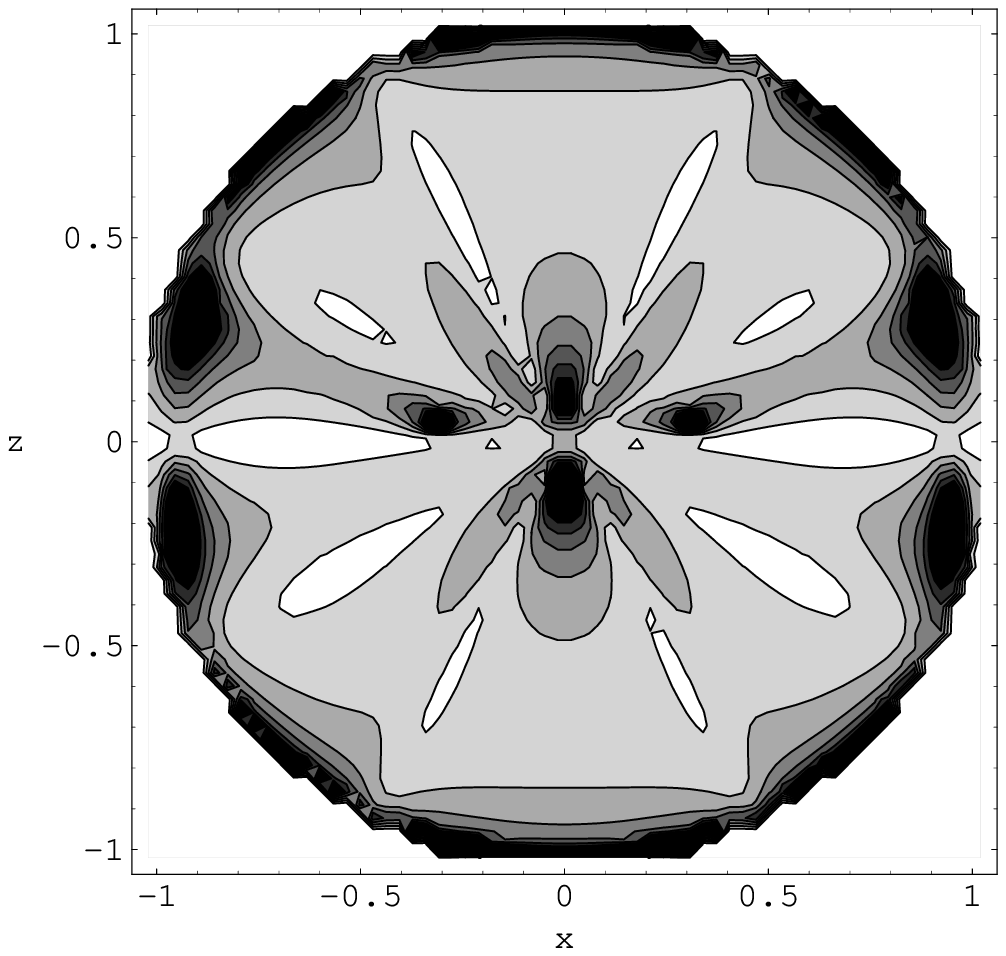}
\end{center}
\caption{Contour plot of the potential $q$ in experiment  2.}
\end{figure}

\section{Conclusions}\label{S:6}
A method is developed for finding the number $N(x)$ of small 
acoustically
soft particles to be embedded per unit volume around every
point $x$ in a bounded domain, filled with a known material,
in order that the resulting new material has the desired
radiation pattern. Any wave field, not
necessarily acoustic wave field, which satifies equations
(9)-(11) is covered by our theory.

On the boundary of each acoustically soft 
particle the Dirichlet condition holds. 

The method is justified theoretically. Numerical 
examples of its application are presented.
The ill-posedness of our problem is discussed and a 
regularization method for its stable solution is proposed
and successfully tested numerically.

The direct application of
the derived formula \eqref{e27} may lead to large
values of $q$. To remedy this situation the coefficients
$h_{l,m}$ are bounded. This is a way to handle the 
ill-posedness of the inverse problem. The resulting 
algorithm exhibits 
a stable behavior. It serves as a regularizing algorithm
for solving the 
original ill-posed problem. Numerical results show that the 
method
can produce materials with the desired focusing properties
under the limitation that the desired radiation pattern
$f(\beta)$ is well approximated by a short series of
spherical harmonics.

\section*{References}

Bateman, H.,  Erdelyi, A. (1954) {\it Tables of integral
transforms},
McGraw-Hill, New York.

Marchenko, V. and Khruslov, E. (1974)
{\it Boundary-value problems in domains with fine-grained boundary},
Naukova Dumka, Kiev, (in Russian).


Ramm, A.G. (1992) {\it Multidimensional inverse scattering problems},
 Longman/Wiley, New York.

Ramm, A.G. (2002)
'Stability of solutions to inverse scattering problems with fixed-energy
data', {\it Milan Journ of Math.}, 70, 97-161.

Ramm, A.G. (2005a)
{\it Inverse problems},  Springer, New York.

Ramm, A.G. (2005b)
{\it Wave scattering by small bodies of arbitrary shapes},
 World Sci. Publishers, Singapore.

Ramm, A.G. (2006a)
'Distribution of particles which produces a ``smart" material',
{\it paper http://arxiv.org/abs/math-ph/0606023}.

Ramm, A.G. (2006b)
 'Distribution of particles which produces a
desired radiation pattern', {\it Communic. in Nonlinear Sci. and Numer.
Simulation}, (to appear).

http://arxiv.org/abs/math-ph/0507006

Ramm, A.G. (2006c)
'Inverse scattering problem with data at fixed energy and
fixed incident direction', submitted.

\end{document}